\documentstyle[12pt,psfig]{article}
\catcode`\@=11
\def\marginnote#1{}
\newcount\hour
\newcount\minute
\newtoks\amorpm
\hour=\time\divide\hour by60
\minute=\time{\multiply\hour by60 \global\advance\minute by-
\hour}
\edef\standardtime{{\ifnum\hour<12 \global\amorpm={am}%
    \else\global\amorpm={pm}\advance\hour by-12 \fi
    \ifnum\hour=0 \hour=12 \fi
    \number\hour:\ifnum\minute<100\fi\number\minute\the\amorpm}}
\edef\militarytime{\number\hour:\ifnum\minute<100\fi\number\minute}
\def\draftlabel#1{{\@bsphack\if@filesw {\let\thepage\relax
  \xdef\@gtempa{\write\@auxout{\string
    \newlabel{#1}{{\@currentlabel}{\thepage}}}}}\@gtempa
    \if@nobreak \ifvmode\nobreak\fi\fi\fi\@esphack}
     \gdef\@eqnlabel{#1}}
\def\@eqnlabel{}
\def\@vacuum{}
\def\draftmarginnote#1{\marginpar{\raggedright\scriptsize\tt#1}}
\def\draft{\oddsidemargin -.5truein
        \def\@oddfoot{\sl preliminary draft \hfil
        \rm\thepage\hfil\sl\today\quad\militarytime}
        \let\@evenfoot\@oddfoot \overfullrule 3pt
        \let\label=\draftlabel
        \let\marginnote=\draftmarginnote

\def\@eqnnum{(\theequation)\rlap{\kern\marginparsep\tt\@eqnlabel}%
\global\let\@eqnlabel\@vacuum}  }
\def\preprint{\twocolumn\sloppy\flushbottom\parindent 1em
        \leftmargini 2em\leftmarginv .5em\leftmarginvi .5em
        \oddsidemargin -.5in    \evensidemargin -.5in
        \columnsep 15mm \footheight 0pt
        \textwidth 250mmin      \topmargin  -.4in
        \headheight 12pt \topskip .4in
        \textheight 175mm
        \footskip 0pt

\def\@oddhead{\thepage\hfil\addtocounter{page}{1}\thepage}
        \let\@evenhead\@oddhead \def\@oddfoot{} \def\@evenfoot{}
}
\def\titlepage{\@restonecolfalse\if@twocolumn\@restonecoltrue\onecolumn
     \else \newpage \fi \thispagestyle{empty}\c@page\z@
        \def\thefootnote{\fnsymbol{footnote}} }
\def\endtitlepage{\if@restonecol\twocolumn \else  \fi
        \def\thefootnote{\arabic{footnote}}
        \setcounter{footnote}{0}}  
\catcode`@=12
\relax
\def\be{\begin{equation}}
\def\ee{\end{equation}}
\def\bea{\begin{eqnarray}}
\def\eea{\end{eqnarray}}
\def\bear{\be\begin{array}}
\def\eear{\end{array}\ee}

\def\NPB#1#2#3{{\it Nucl.~Phys.} {\bf{B#1}} (19#2) #3}
\def\PLB#1#2#3{{\it Phys.~Lett.} {\bf{B#1}} (19#2) #3}
\def\PRD#1#2#3{{\it Phys.~Rev.} {\bf{D#1}} (19#2) #3}
\def\PRL#1#2#3{{\it Phys.~Rev.~Lett.} {\bf{#1}} (19#2) #3}

\def\mst11{m_{\;\widetilde{t}_{1}}}

\def\mst22{m_{\;\widetilde{t}_{2}}}
\def\mst12{m_{\;\widetilde{t}_{1,2}}}

\def\msb11{m_{\;\widetilde{b}_{1}}}
\def\msb22{m_{\;\widetilde{b}_{2}}}
\def\msb12{m_{\;\widetilde{b}_{1,2}}}

\def\mtilde2{\widetilde{m}^{2}}

\relax

\def\ula{u_{L1} (r, \theta)}
\def\ulb{u_{L2} (r, \theta)}
\def\ura{u_{R1} (r, \theta)}
\def\urb{u_{R2} (r, \theta)}

\def\dla{d_{L1} (r, \theta)}
\def\dlb{d_{L2} (r, \theta)}
\def\dra{d_{R1} (r, \theta)}
\def\drb{d_{R2} (r, \theta)}

\begin{document}

\topmargin-2.5cm

\begin{titlepage}
\begin{flushright}
IEM-FT-111/95 \\
hep--ph/9508404 \\
\end{flushright}
\vskip 0.3in
\begin{center}{\Large\bf
Top-bottom doublet in the sphaleron background~\footnote{Work
supported in part by
the European Union (contract CHRX-CT92-0004) and
CICYT of Spain
(contract AEN94-0928).}  } \\
\vskip .5in
{\large {\bf J. Moreno, D.H. Oaknin}\\
\vspace{.5cm}
Instituto de Estructura de la Materia, CSIC, Serrano 123,
28006-Madrid, Spain\\
and\\
\vspace{.5cm}
{\bf M. Quir\'os}~\footnote{On leave of absence from Instituto
de Estructura de la Materia, CSIC, Serrano 123, 28006-Madrid,
Spain.}\\
\vspace{.5cm}
CERN, TH Division, CH--1211 Geneva 23, Switzerland }\\
\vskip.35in
\end{center}
\vskip1.cm
\begin{center}
{\bf Abstract}
\end{center}
\begin{quote}
We consider the top-bottom doublet in the background of the sphaleron
for the realistic case of large non-degeneracy of fermion masses,
in particular $m_b=5$ GeV and $m_t=175$ GeV.
We propose an axially symmetric $(r,\theta)$-dependent ansatz
for fermion fields and investigate the effects of the non-degeneracy
on them. The exact solution is described, with an error less than 0.01\%,
by a set of ten radial functions. We also propose an approximate
solution, in the $m_b/m_t\rightarrow 0$ limit,
with an error ${\cal O}(m_b/m_t)$. We have found that the effects of
non-degeneracy provide a $\theta$-dependence typically $\sim 10\%$.

\end{quote}
\vskip2.5cm

\begin{flushleft}
August 1995 \\
\end{flushleft}

\end{titlepage}

\setcounter{footnote}{0}
\setcounter{page}{0}
\newpage
%

After the discovery by t'Hooft~\cite{tH} that baryon and lepton
numbers are anomalous global symmetries in the Standard Model, violated
by non-perturbative effects, and the observation~\cite{KRS} that the
three Sakharov's conditions for baryogenesis can be fulfilled at the
electroweak phase transition, there has been renewed interest in
understanding the dynamics of baryon asymmetry generation
near the electroweak critical temperature. In particular the sphaleron,
a static unstable solution of the classical equations of motion~\cite{MKM},
has played a key r\^ole because the baryon violation rate for
fluctuations between neighboring minima contains a Boltzmann
suppression factor proportional to the energy of the sphaleron
$E_{\rm sph}$. In particular the rate per unit time and unit volume
is given by~\cite{AML}
\be
\label{rate}
\Gamma\sim\omega T^4\exp\left(-\frac{E_{\rm sph}}{v}
\frac{\phi(T)}{T}\right)
\ee
where the sphaleron solution is usually computed in the limit of
zero Weinberg angle (corrections of ${\cal O}(g')$ have been proved
to be negligible~\cite{KKB} in the absence of fermions)
and use has been made that the energy of
the sphaleron at finite temperature follows an approximate scaling law
with $\phi(T)/v$~\cite{BK}. The prefactor $\omega$ in (\ref{rate})
which contains, in particular,
the product of bosonic and fermionic fluctuation
determinants ($\kappa_{\rm bos}\kappa_{\rm fer}$) around the sphaleron,
was evaluated~\cite{omega} in the high $T$ limit
(three-dimensional effective theory
with all fermionic modes decoupled, i.e. $\kappa_{\rm fer}=1$).

In view of the recent experimental detection of the top-quark and its
mass measurement~\cite{top} it seems of utmost importance to consider
the modification, due to the presence of the top-quark, of the classical
sphaleron energy and its contribution to $\kappa_{\rm fer}$. Progress in
this direction has been done by considering both effects, in
Refs.~\cite{Kunz} and \cite{Diakonov},
respectively, as triggered by the presence of
mass degenerate fermion doublets. The reason for studying degenerate
fermions is simplicity since the spherical
symmetry of the sphaleron solution
is not spoiled by the presence of degenerate fermion doublets.

However, for the top-bottom doublet, with hierarchically different masses,
$m_b\sim 5$ GeV and $m_t\sim 175$ GeV, the breaking of degeneracy
(custodial symmetry), controlled by $(m_t-m_b)/m_t$, cannot be
considered {\it a priori} as a small perturbation and the previous
results might be spoiled by the large non-degeneracy effects.
It is the purpose of this note to study the effects of the
non-degeneracy of the top-bottom doublet in the background of the
sphaleron. Our results have to be considered as a first step towards
quantifying the effects of the non-degeneracy
of fermion doublets on the sphaleron energy. They contain
interesting (and somewhat unexpected) results which could be used as
hints towards computing the effects of the top-bottom doublet with physical
masses on the classical sphaleron barrier and fermion determinant.

After completion of this work we have received Ref.~\cite{NKK} where
the same problem is considered using a different expansion for the
fermion fields. We will compare both approaches and point out the
differences in both expansions.

We will now study the fermionic distribution of a realistic quark doublet,
non-degenerate in mass, in the sphaleron background. To simplify, we will
work in the $g' = 0$ approximation. In this limit the hypercharge field
$B_\mu$ decouples and the sphaleron energy density is spherically symmetric.
The sphaleron configuration can be parametrized as:

\bear{lcl}
\label{sphal}
\Phi(\bf \hat r) & = &  i   {\hat r}. {\vec \tau} h(r)
\left[\begin{array}{r}   0  \\   1  \end{array}\right],  \\
& & \\
W^a_i & = &  \frac{2}{gr} \epsilon_{aij} {\hat r}_j f(r)
\eear
\vspace{.25cm}

The relevant lagrangian density for fermions is:
\bear{ccl}
\label{lagrangian}
  {\cal L} & = &  \phantom{+}  \; i
{\overline {\bf Q}} {\not \! \! D} {\bf Q} \;
        +  i {\overline {\bf u}_R} {\not \! \partial} {\bf u}_R\;
   +     i {\overline {\bf d}_R} {\not \! \partial} {\bf d}_R\; \\
& & \\
  &   &  -  h_t \left( {\overline  {\bf Q}} {\tilde \Phi} {\bf u}_R
                               + h.c. \right)
 -  h_b \left( {\overline  {\bf Q}} \Phi {\bf d}_R + h.c. \right)
\eear
where
$$ {\bf Q}=\left(
\begin{array}{c}
{\bf u}_L \\
{\bf d}_L
\end{array} \right)
$$
is the left-handed quark doublet,
$W_{\mu \nu}^a = \partial_{\mu} W_{\nu}^a -
                 \partial_{\nu} W_{\mu}^a  +
                 g \varepsilon^{abc} W_{\mu}^b W_{\nu}^c$,
$ D_{\mu} = \partial_{\mu} - i \frac{g}{2} \tau^a W_{\mu}^a $,
${\tilde \Phi} \equiv i \tau^2 \Phi^*$.

The equations of motion for the fermionic fields can be written as:

\bear{ccl}
\label{eqnold}
i {\not \! \! D}  {\bf Q}        & = &  h_t {\tilde \Phi} {\bf u}_R +
                                        h_b         \Phi  {\bf d}_R  \\
& & \\
i {\not \!  \partial} {\bf u}_R  & = &  h_t {\tilde \Phi}^{\dag}  {\bf Q} \\
& & \\
i {\not \!  \partial} {\bf d}_R  & = &  h_b \Phi^{\dag}  {\bf Q} \\
\eear
The non degeneracy of fermions explicitly breaks the global
custodial symmetry $SU(2)_L \times SU(2)_R$. Let us remember that
in the degenerate case the Dirac hamiltonian deduced from
(\ref{eqnold}) commutes with the grand spin operator
$\vec{K}=\vec{L}+\vec{S}+\vec{T}$, where $\vec{L}$ stands for the
angular momentum, $\vec{S}$ for the spin and $\vec{T}$ for the
isospin. In this case the eigenstates $\Psi$ of the Dirac operator can be
labelled by $k,k_3$, with $K^2\Psi= k(k+1)\Psi$
and $K_3\Psi = k_3\Psi$. In particular the
zero mode is included in the $k=0$ subspace. For non-degenerate fermions,
the different Yukawa couplings to the Higgs field background
single out a direction in the isospin space. As a consequence
$K^2$ does not commute any longer with the hamiltonian and
the energy eigenstates will have non-vanishing projections over
different values of $k$.
However, $K_3$ still commutes with the hamiltonian, and hence $k_3$ is a
good quantum number to label the energy eigenstates. Since $k_3$ is a
discrete label and we want to continuously deform the
zero mode from the degenerate case to the non-degenerate one,
we will restrict ourselves to the subspace $k_3=0$.

To describe the fermions, we deal with the Dirac representation:
\be
\label{dirac}
\gamma^0 = \left( \begin{array}{cr} \sigma^0&     0      \\
                                        0  &  -\sigma^0 \end{array}\right) \;\;
\gamma^i = \left( \begin{array}{rc}     0  & -\sigma^i  \\
                                  \sigma^i &   0       \end{array}\right) \;\;
\gamma^5 = \left( \begin{array}{cc}     0  & {\bf 1}   \\
                                 {\bf 1}   &  0        \end{array}\right)
\ee
with $\sigma^i$ the Pauli matrices and $\sigma^0 = {\bf 1}$.
Using two component spinors, $q_L$, $q_R$, defined as:
\be
\label{twospinors}
{\bf q}_L = \frac{1}{\sqrt{ 2} }
                \left[\begin{array}{r}   q_L  \\  - q_L
                      \end{array}\right],  \; \; \;
{\bf q}_R = \frac{1}{\sqrt{ 2} }
                \left[\begin{array}{r}   q_R  \\    q_R
                      \end{array}\right],  \; \; \;
\ee
we can write our ansatz, a general parametrization of the $k_3=0$
subspace, as:

\bear{lcl}
u_L({\vec r}) & = & ( u_{L2}(\rho,z)   +
             u_{L1}(\rho,z) \, {\bf {\hat \rho}}.{\vec \sigma})
\,\chi_{-} \\ \\
d_L({\vec r}) & = & ( d_{L1}(\rho,z)   +
             d_{L2}(\rho,z) \, {\bf {\hat \rho}}.{\vec \sigma})
\,\chi_{+} \\ \\
u_R({\vec r}) & = & ( u_{R2}(\rho,z)   +
             u_{R1}(\rho,z) \, {\bf {\hat \rho}}.{\vec \sigma})
\,\chi_{-} \\ \\
d_R({\vec r}) & = & ( d_{R1}(\rho,z)   +
             d_{R2}(\rho,z) \, {\bf {\hat \rho}}.{\vec \sigma}) \,\chi_{+}
\eear
where we have decomposed ${\vec r}$ into the longitudinal and transverse
components,
$
        {\vec r} = {\vec \rho} + z {\vec k},
$
$\hat {\bf  \rho} $ is the normalized vector in the XY plane
$     {\vec \rho} / \rho $,
$\hat {\bf \rho} . {\vec  \sigma} = \rho_x \sigma_x  + \rho_y \sigma_y$
and
\be
 \chi_{+} = \left[\begin{array}{r}   1  \\   0  \end{array}\right],  \; \; \;
 \chi_{-} = \left[\begin{array}{r}   0  \\   1  \end{array}\right]
\ee

The equations for fermions can be written as  functions of
$r,\theta$, with $\rho = r \sin \theta$, as:


$$
\begin{array}{ll}
\partial_r \ula + \frac{1}{r} \ula  +  \frac{1}{r} \partial_\theta \ulb & \\
+ \frac{1}{r} f(r) \left( \frac{1}{2} \sin{2\theta} (\dla + \ulb)
                          + \sin^2\theta(\dlb - \ula)           \right) &  \\
-h(r) \left( \frac{1}{2} \sin{2\theta} (h_b \dra + h_t \urb)
 + \cos^2\theta h_t \ura  + \sin^2\theta h_b \drb \right) =  &  0    \\ \\
\partial_r \ulb - \cot\theta \frac{1}{r} \ula
                - \frac{1}{r} \partial_\theta \ula & \\
- \frac{1}{r} f(r) \left(  (\dla - \ulb) + \cos^2\theta(\dla + \ulb)\right. &
\\
\phantom{- \frac{1}{r} f(r)}\left.  + \frac{1}{2} \sin{2\theta} (\dlb - \ula)
\right) &  \\
- h(r) \left( \frac{1}{2} \sin{2\theta} (h_t \ura - h_b \drb)
 + \sin^2\theta h_b \dra  - \cos^2\theta h_t \urb \right)  = &  0    \\ \\
\partial_r \dla + \cot\theta \frac{1}{r} \dlb
                + \frac{1}{r} \partial_\theta \dlb & \\
- \frac{1}{r} f(r) \left(  (\ulb - \dla) + \cos^2\theta(\dla + \ulb)
\right.  & \\
\phantom{- \frac{1}{r} f(r)}  \left.
              + \frac{1}{2} \sin{2\theta} (\dlb - \ula)       \right) &  \\
- h(r) \left( \frac{1}{2} \sin{2\theta} (h_t \ura - h_b \drb)
 + \sin^2\theta h_t \urb  - \cos^2\theta h_b \dra \right)  = &  0    \\ \\
\partial_r \dlb + \frac{1}{r} \dlb  -  \frac{1}{r} \partial_\theta \dla & \\
- \frac{1}{r} f(r) \left( \frac{1}{2} \sin{2\theta} (\dla + \ulb)
                          + \sin^2\theta(\dlb - \ula)           \right) &  \\
+h(r) \left( \frac{1}{2} \sin{2\theta} (h_t \urb + h_b \dra)
 - \sin^2\theta h_t \ura  - \cos^2\theta h_b \drb \right)  = &  0    \\ \\
%
%
\partial_r \ura + \frac{1}{r} \ura  +  \frac{1}{r} \partial_\theta \urb & \\
-h(r) h_t \left( \frac{1}{2} \sin{2\theta} ( \dla +  \ulb)
 + \cos^2\theta  \ula  + \sin^2\theta  \dlb \right)  = &  0    \\ \\
\partial_r \urb - \cot\theta \frac{1}{r} \ura
                - \frac{1}{r} \partial_\theta \ura & \\
- h(r) h_t \left( \frac{1}{2} \sin{2\theta} ( \ula - \dlb)
 + \sin^2\theta  \dla  - \cos^2\theta  \ulb \right)  = &  0    \\ \\
\end{array}
$$
\begin{equation}
\label{eqferm}
\begin{array}{ll}
\partial_r \dra + \cot\theta \frac{1}{r} \drb
                + \frac{1}{r} \partial_\theta \drb & \\
- h(r) h_b \left( \frac{1}{2} \sin{2\theta} ( \ula - \dlb)
 + \sin^2\theta  \ulb  - \cos^2\theta \dla \right)  = &  0    \\ \\
\partial_r \drb + \frac{1}{r} \drb  -  \frac{1}{r} \partial_\theta \dra & \\
+h(r) h_b \left( \frac{1}{2} \sin{2\theta} (\ulb + \dla)
 - \sin^2\theta \ula  - \cos^2\theta  \dlb \right)  = &  0    \\ \\
\end{array}
\end{equation}

We have, then, a set of eight partial differential equations.
Instead of solving it in general, we will work out a
systematic expansion of these functions in terms of a Fourier
series. Notice that the commutator of the hamiltonian
[defined on the $(q_L,q_R)^T$-space] with parity vanishes on the zero modes,
which allows to label them with a definite parity. As parity is a discrete
symmetry we will choose for our ansatz the
same parity (i.e. even) as zero modes in the degenerate case.
Therefore, our expansion can be written
as~\footnote{An alternative expansion is given by developing
the fermionic fields in terms of
the even spherical harmonic subsets
$Y_{2n}^0 (\theta)$, $Y_{2n}^{\pm 1} (\theta,\phi) $
instead of $\cos(2n\theta)$, $\sin(2n\theta)e^{\pm i \phi}$
respectively. In this
basis it is straightforward to check that $k_3=0$.}

\bear{rclcr}
\label{Fourier}
\ula & = &                & &\sum_{n=1}^{\infty}u_{L1}^{(n)}(r)
                             \sin(2n\theta)\\\\
\ulb & = & u_{L2}^{(0)}(r)&+&\sum_{n=1}^{\infty}u_{L2}^{(n)}(r)
                             \cos(2n\theta)\\\\
\dla & = & d_{L1}^{(0)}(r)&+&\sum_{n=1}^{\infty}d_{L1}^{(n)}(r)
                             \cos(2n\theta)\\\\
\dlb & = &                & &\sum_{n=1}^{\infty}d_{L2}^{(n)}(r)
                             \sin(2n\theta)\\\\
\ura & = &                & &\sum_{n=1}^{\infty}u_{R1}^{(n)}(r)
                             \sin(2n\theta)\\\\
\urb & = & u_{R2}^{(0)}(r)&+&\sum_{n=1}^{\infty}u_{R2}^{(n)}(r)
                              \cos(2n\theta)\\\\
\dra & = & d_{R1}^{(0)}(r)&+&\sum_{n=1}^{\infty}d_{R1}^{(n)}(r)
                             \cos(2n\theta)\\\\
\drb & = &                & &\sum_{n=1}^{\infty}d_{R2}^{(n)}(r)
                             \sin(2n\theta)
\eear

We expect the convergence of the series (\ref{Fourier}) to be
good enough to provide a reliable solution to the system of equations
(\ref{eqferm}). In fact, cutting off the series at $n=0$ would
include the appropriate ansatz to describe a situation in which the fermion
doublet is degenerate. However the case we want to address, i.e. that
of the top-bottom doublet, is very massive and highly non-degenerate.
This means, on the one hand, that we need the modes $n\geq 1$ in
(\ref{Fourier}) and, on the other hand, that the convergence of the
series (\ref{Fourier}) might be slow enough to require
many terms for an accurate description of the solution.
Fortunately we will see this is not the case, and, as we will prove,
taking only the terms with $n=0,1$ provides a reliable solution with an
accuracy better than 0.01\%.

Assuming the convergence of series (\ref{Fourier}) one can solve the system
(\ref{eqferm}) to any given order $n$ (i.e. neglecting all modes with
higher values of $n$) and estimate the accuracy of the different
approximations. A first step along this direction is solving
(\ref{eqferm}) to order $n=0$, i.e. in the approximation of a
spherically symmetric ansatz for fermions (as in the case $m_u=m_d$).
Only four spinor components do not vanish:
\begin{equation}
\label{setzero}
u_{L2}^{(0)},\ d_{L1}^{(0)},\ u_{R2}^{(0)},\ d_{R1}^{(0)}.
\end{equation}
Replacing them
into (\ref{eqferm}) one obtains a system of four differential
equations whose solution is shown in Fig.~1 (dashed lines)~\footnote{All
dimensional quantities are expressed in units of the corresponding power
of $v=\langle\phi_0\rangle$, and we have arbitrarily normalized the total
fermionic density to $4\pi$.}. It is interesting to notice that the
relation
\begin{equation}
\label{leftzero}
u_{L2}^{(0)}(r)+d_{L1}^{(0)}(r)=0,
\end{equation}
valid in the degenerate case, also holds in this approximation.
On the other hand, the similar relation valid in the degenerate case,
$u_{R2}^{(0)}(r)+d_{R1}^{(0)}(r)=0$ is spoiled by non-degeneration effects,
and replaced by
\begin{equation}
\label{rightzero}
h_t d_{R1}^{(0)}(r)+h_b u_{R2}^{(0)}(r)=0,
\end{equation}
In fact, the functions defined by the left-hand sides of
Eqs.~(\ref{leftzero}) and (\ref{rightzero})
only couple to themselves and to higher modes, and hence can be consistently
fixed to zero to this order.

In general, solving the system to a given order $n$, amounts to keeping
the system of fields
\begin{eqnarray}
\label{setn}
u_{L1}^{(p)},\ u_{L2}^{(p)},\ d_{L1}^{(p)},\  d_{L2}^{(p)},\nonumber \\
\nonumber \\
u_{R1}^{(p)},\ u_{R2}^{(p)},\ d_{R1}^{(p)},\  d_{R2}^{(p)}
\end{eqnarray}
for $p=1,...,n$, and the spherically symmetric components (\ref{setzero}):
a system of $8n+4$ differential equations with $8n+4$ unknown functions.
We will solve the system of differential equations using the following
boundary conditions. There are two zeroes in the $(8n+4)\times(8n+4)$
asymptotic mass matrix which correspond to two directions, in the space of
highest, $n$-th, components of the spinorial
fields. They must be fixed to zero
at infinity, and thus elsewhere since these directions
have (up to higher order contributions) fixed points at zero.
Fixing to zero these two directions amounts to the conditions:
\begin{equation}
\label{leftn}
u_{L1}^{(n)}(r)+u_{L2}^{(n)}(r)+d_{L1}^{(n)}(r)-d_{L2}^{(n)}(r)=0
\end{equation}
and
\begin{equation}
\label{rightn}
h_b u_{R1}^{(n)}(r)+h_b u_{R2}^{(n)}(r)+
h_t d_{R1}^{(n)}(r)-h_t d_{R2}^{(n)}(r)=0
\end{equation}
Out of the remaining $8n+2$ fields,
$4n+1$ correspond to positive mass eigenvalues, and therefore are fixed
to zero at infinity imposing normalizability of the fermionic
configuration,
and $4n+1$ have negative mass eigenvalues. $4n$ of
them are fixed to zero at the origin (triggering the vanishing at the origin
of all $8n$ $\theta$-dependent components), and the last one
is used to fix the global normalization.
In particular, for $n=1$ we obtain a system of twelve differential equations
with twelve unknown functions: the set (\ref{setzero}) and (\ref{setn}) with
$p=1$. The solution is presented in Figs.~1 and 2 (solid lines). We can
see from Fig.~1 that the relations (\ref{leftzero}) and (\ref{rightzero})
are spoiled by the presence in the equations of the angular dependent
components. The spoiling is $\sim$10\%, which is the expected order of
magnitude of the angular dependent part of the solution.
The latter, i.e. the radial functions of (\ref{setn})
with $p=1$ are plotted in Fig.~2 (solid lines). We can see that the typical
size of these functions is $\sim 10$\% those contributing to the
spherically symmetric part of Fig.~1. This is the amount by which the
relations between the components with $n=0$, Eqs.~(\ref{leftzero}) and
(\ref{rightzero}), are spoiled. The latter are replaced by the similar
relations (\ref{leftn}) and (\ref{rightn}) between the $n=1$ components.
One can easily check from the curves in Fig.~2 that relations
(\ref{leftn}) and (\ref{rightn}) are exactly satisfied.
Moreover, we have numerically found that Eq.~(\ref{leftn})
splits with a good accuracy,
for $n\geq 1$, into the couple of constraints:
\begin{equation}
\label{leftn1}
u_{L1}^{(n)}(r)+d_{L1}^{(n)}(r)=0
\end{equation}
and
\begin{equation}
\label{leftn2}
u_{L2}^{(n)}(r)-d_{L2}^{(n)}(r)=0 .
\end{equation}
We believe that Eqs.~(\ref{leftn1}) and (\ref{leftn2}) are not
a numerical coincidence. The fermionic fields expanded up to
a given order $n$ can be decomposed into
eigenstates of angular momentum $\ell=0,\; 2,\; ,\dots,\; \leq 2n$,
isospin $T_3=\pm 1/2$ and spin $S_3=\pm 1/2$, given by
$$
\left| \ell ,-(T_3+S_3)\right\rangle \otimes
\left| \frac{1}{2}, T_3\right\rangle \otimes
\left| \frac{1}{2}, S_3
\right\rangle .
$$
Defining $\vec{J}=\vec{L}+\vec{T}$ and using the basis
of $J^2,J_3$ eigenstates, condition (\ref{leftn1})
is equivalent to projecting out
the left-handed states on
$\left| j,j_3 \right\rangle=
\left| 2n+\frac{1}{2},-\frac{1}{2} \right\rangle$.
Similarly, condition (\ref{leftn2}) amounts to
projecting out on
$\left| j,j_3 \right\rangle =
\left| 2n-\frac{1}{2},\frac{1}{2}\right \rangle $.
Therefore a consistent ansatz using the latter basis should
take advantage of this approximate symmetry.

We have solved the system (\ref{eqferm}) to order $n=2$. In this case
we have a system of twenty differential
equations with twenty unknown functions,
those in (\ref{setn}) with $p=1,2$ and in (\ref{setzero}). The new
components, with $p=2$ in (\ref{setn}), are less than $10^{-4}$ the
large components in Fig.~1 while the $p=1$ components in (\ref{setn})
and the components in (\ref{setzero}), computed to this order of
approximation, are indistinguishable from those previously computed.
In fact, we have plotted in Figs.~1 and 2 the latter in solid lines and
they are superimposed on those computed to order $n=1$. In the same way
the relationships (\ref{leftn1}) and (\ref{leftn2})
remain very accurately satisfied.

\newpage
To summarize the previous results, we can keep, out of the general
expansion (\ref{Fourier}), the particular reduction
\begin{equation}
\label{order1}
\begin{array}{ll}
u_L(\vec{r})=\left[
\begin{array}{rcl}
                & & u_{L1}^{(1)}(r)\sin 2\theta e^{-i\phi} \\ && \\
u_{L2}^{(0)}(r) &+& u_{L2}^{(1)}(r)\cos 2\theta
\end{array}
\right]
&
u_R(\vec{r})=\left[
\begin{array}{rcl}
                & & u_{R1}^{(1)}(r)\sin 2\theta e^{-i\phi} \\&& \\
u_{R2}^{(0)}(r) &+& u_{R2}^{(1)}(r)\cos 2\theta
\end{array}
\right]
\\
& \\
d_L(\vec{r})=\left[
\begin{array}{rcl}
d_{L1}^{(0)}(r) &+& d_{L1}^{(1)}(r)\cos 2\theta \\&& \\
                & & d_{L2}^{(1)}(r)\sin 2\theta e^{i\phi}
\end{array}
\right]
&
d_R(\vec{r})=\left[
\begin{array}{rcl}
d_{R1}^{(0)}(r) &+&{\displaystyle
 d_{R1}^{(1)}(r)\cos 2\theta} \\&& \\
                & & {\displaystyle
 d_{R2}^{(1)}(r)\sin 2\theta e^{i\phi} }
\end{array}
\right]\\
\end{array}
\end{equation}
where two radial functions can be eliminated using the relations
(\ref{leftn}) and (\ref{rightn}).

Given the smallness of the ratio $m_b/m_t$ one can try to implement
the $m_b/m_t\rightarrow 0$ limit in our solution. In that limit $d_R$
decouples from the equations of motion and can be consistently fixed to
zero. In particular
\begin{equation}
\label{rightnap}
d_{R1}^{(n)}(r)=d_{R2}^{(n)}(r)\equiv 0.
\end{equation}
Eq.~(\ref{rightnap}) is very accurately satisfied,
as can be checked from Figs.~1b and 2b.
We have solved the problem for values of $m_b$ in the range
$0\leq m_b\leq 5$ GeV. We have verified in this range that
$d_R$ is proportional to $h_b$.
The rest of fields remain essentially unchanged since
their dependence on $m_b$ is through
${\cal O}(h_b^2)$ terms. Numerically these non-vanishing
fields, computed in the strict limit $m_b=0$,
are superimposed with the corresponding functions plotted
in Figs.~1 and 2 for the physical value of $m_b$.
We can conclude that neglecting all right-handed
components of spinor fields, in the limit $m_b/m_t\rightarrow 0$, is
accurate to ${\cal O}(m_b/m_t)$.

We will compare now the previous solution with that recently
proposed in Ref.~\cite{NKK}.
The solution in Ref.~\cite{NKK} corresponds to the
parametrization in terms of eight radial functions, as:
\begin{equation}
\label{nolte}
\begin{array}{c}
u_L(\vec{r})=\left[
\begin{array}{rl}
  - & \frac{1}{2}F_L^{+}\sin 2\theta e^{-i\phi} \\ & \\
\phantom{-}G_L^{+}+\frac{1}{2}F_L^{+}  +
& \frac{1}{2}F_L^{+}\cos 2\theta
\end{array}
\right]
 \\  \\
u_R(\vec{r})=\left[
\begin{array}{rl}
- & \frac{1}{2}F_R^{+}\sin 2\theta e^{-i\phi} \\ & \\
\phantom{-}G_R^{+}+\frac{1}{2}F_R^{+} +
& \frac{1}{2}F_R^{+}\cos 2\theta
\end{array}
\right]
\\
 \\
d_L(\vec{r})=\left[
\begin{array}{rl}
-G_L^{-} + \frac{1}{2}F_L^{-}+ &
\frac{1}{2}F_L^{-}\cos 2\theta\\ & \\
& \frac{1}{2}F_L^{-}\sin 2\theta e^{i\phi}
\end{array}
\right]
 \\ \\
d_R(\vec{r})=\left[
\begin{array}{rl}
-G_R^{-}+\frac{1}{2}F_R^{-} + &
\frac{1}{2}F_R^{-}\cos 2\theta\\ & \\
& \frac{1}{2}F_R^{-}\sin 2\theta e^{i\phi}
\end{array}
\right]
\end{array}
\end{equation}
where
\begin{equation}
\begin{array}{c}
F_L^{\pm}=F_L(r)\pm\Delta F_L(r), \; \; G_L^{\pm}=G_L(r)\pm\Delta F_L(r), \\
\\
F_R^{\pm}=F_R(r)\pm\Delta F_R(r),\; \; G_R^{\pm}=G_R(r)\pm\Delta F_R(r)
\end{array}
\end{equation}

Since we have proved that our expansion (\ref{order1}) describes the
exact solution with an error less than 0.01\%,
we can compare it with
the parametrization in Eq.~(\ref{nolte}).
In the left-handed sector Eq.~(\ref{nolte}) would
imply that $u_{L1}^{(1)}+u_{L2}^{(1)}=0$ and
$d_{L1}^{(1)}=d_{L2}^{(1)}$: these conditions are consistent with
(\ref{leftn}) though a quick glance at Fig.~2 shows that they
are violated in the exact solution by $\sim$10\% of the
corresponding components, an error much greater than that induced
by the next order. Moreover, these conditions conditions
are not controlled by $m_b/m_t$ but, contrarily, become exact in the
limit of large degeneracy.
Similarly, in the right-handed sector,
Eq.~(\ref{nolte}) would predict $u_{R1}^{(1)}+u_{R2}^{(1)}=0$ and
$d_{R1}^{(1)}=d_{R2}^{(1)}$, which are also consistent with our $n=1$
constraint (\ref{rightn}) but not satisfied by the exact solution.
In summary, we have
found that the parametrization (\ref{nolte}) is accurate in the case
of large degeneracy, but has corrections
not controlled by the parameter $m_b/m_t$, which are
much greater than those corresponding to the next order in our
expansion.

In conclusion, we have performed a consistent expansion for the top-bottom
fermionic fields in the background of the sphaleron. We have found that
the first two terms in the expansion provide the fermionic fields with
an accuracy better than 0.01\%. We have quantified the effect of the
top-bottom non-degeneracy in the $\theta$-dependent part of the fermionic
components as $\sim$10\%. We have proved that the approximation
$m_b/m_t\rightarrow 0$ provides a solution accurate to ${\cal O}(m_b/m_t)$.
Our work should be considered as a first step towards
evaluating the effects of the non-degeneracy of the top-bottom doublet
in the sphaleron energy.

\begin{figure}
\centerline{\vbox{
\psfig{figure=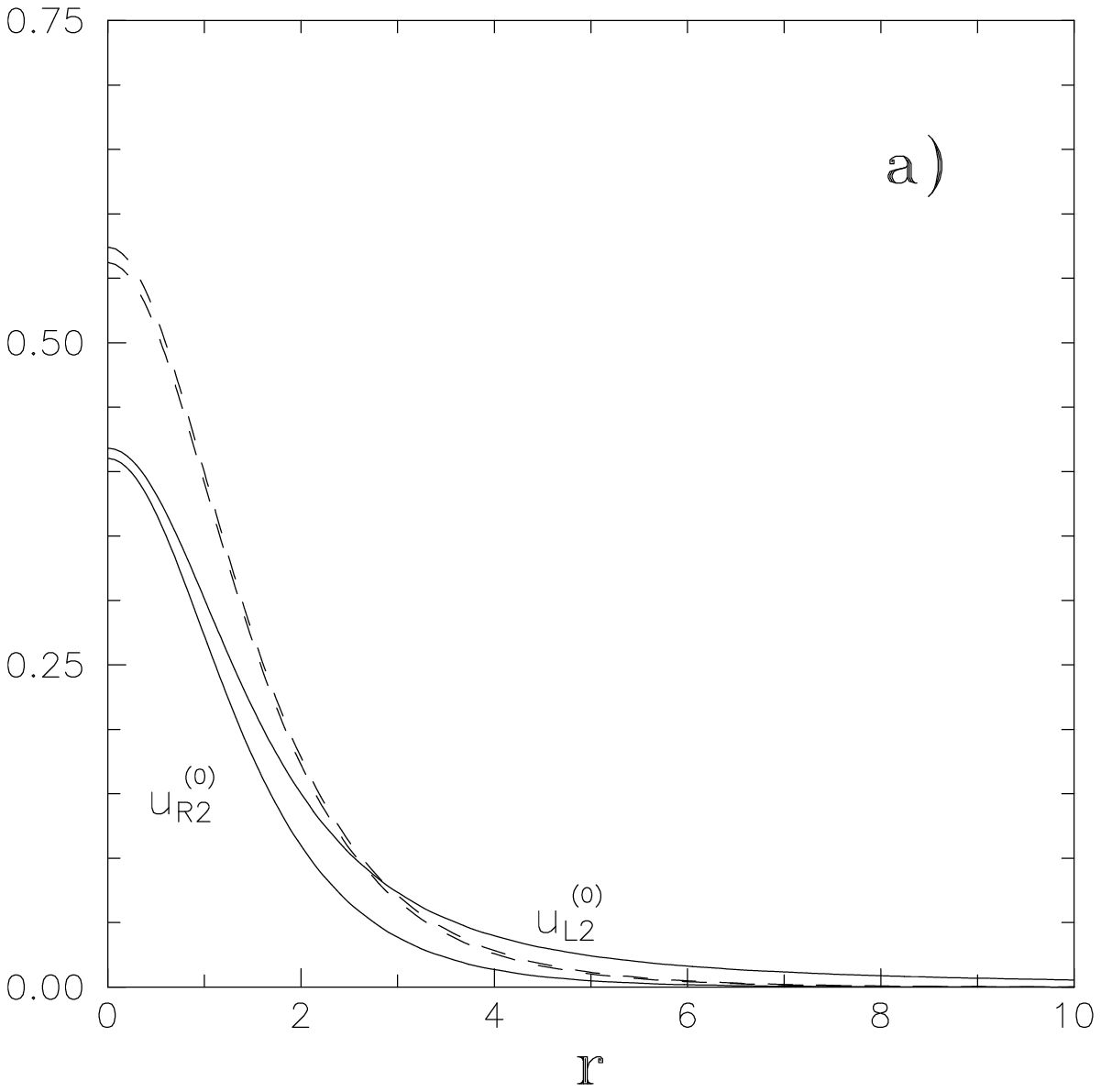,height=9cm,bbllx=4.cm,bblly=2.cm,bburx=14cm,bbury=14cm}
\psfig{figure=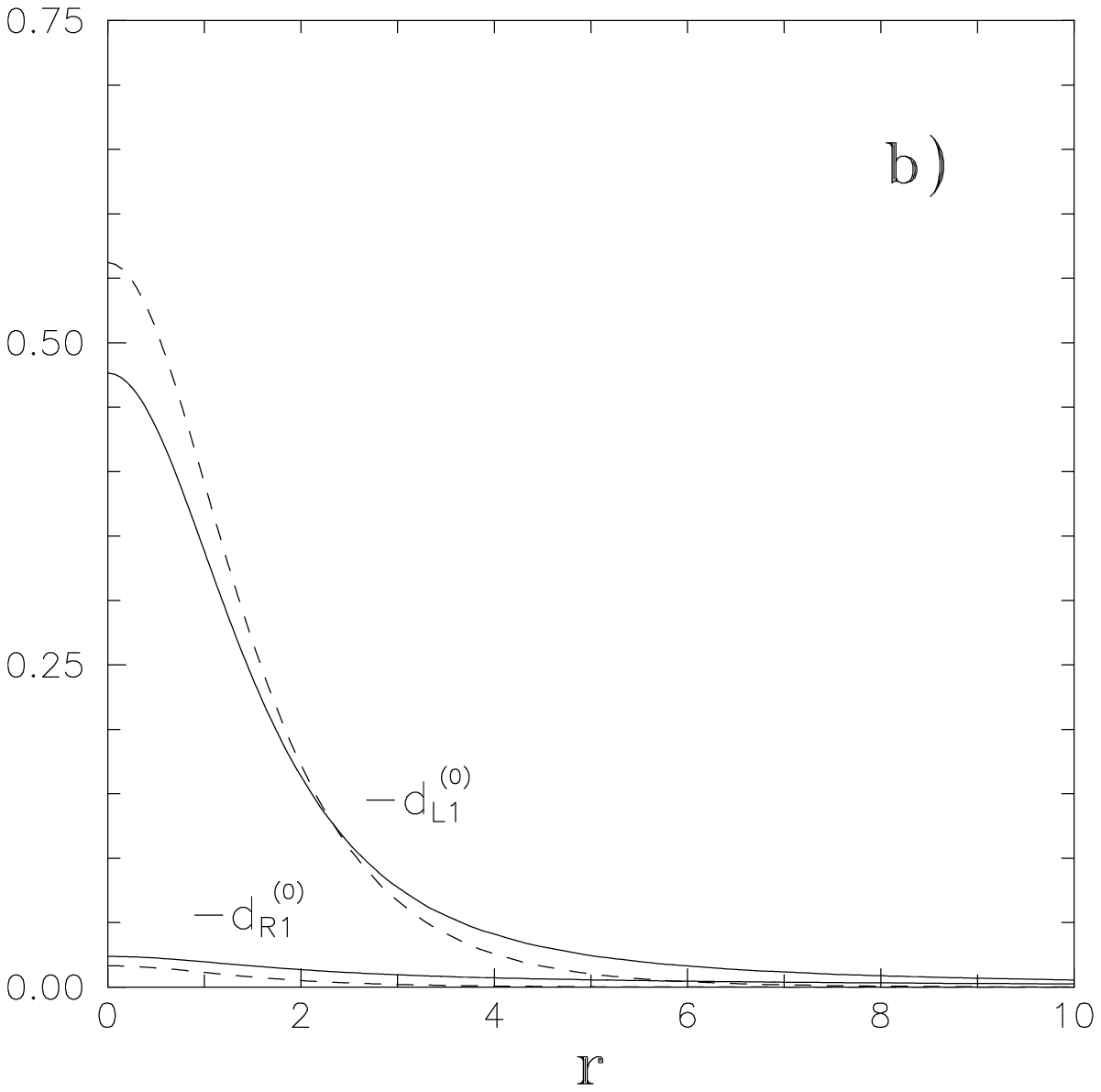,height=9cm,bbllx=4.cm,bblly=2.cm,bburx=14cm,bbury=14cm}
}}
\caption{\tenrm\baselineskip=10pt {\bf a)} The radial components
$u_{L2}^{(0)}$ and $u_{R2}^{(0)}$ computed to order $n=0$ (dashed lines)
and $n=1$ and $2$ (solid lines).
{\bf b)} The radial components $-d_{L1}^{(0)}$ and $-d_{R1}^{(0)}$ computed
to order $n=0$ (dashed lines) and $n=1$ and $2$ (solid lines).}
\end{figure}

\begin{figure}
\centerline{\vbox{
\psfig{figure=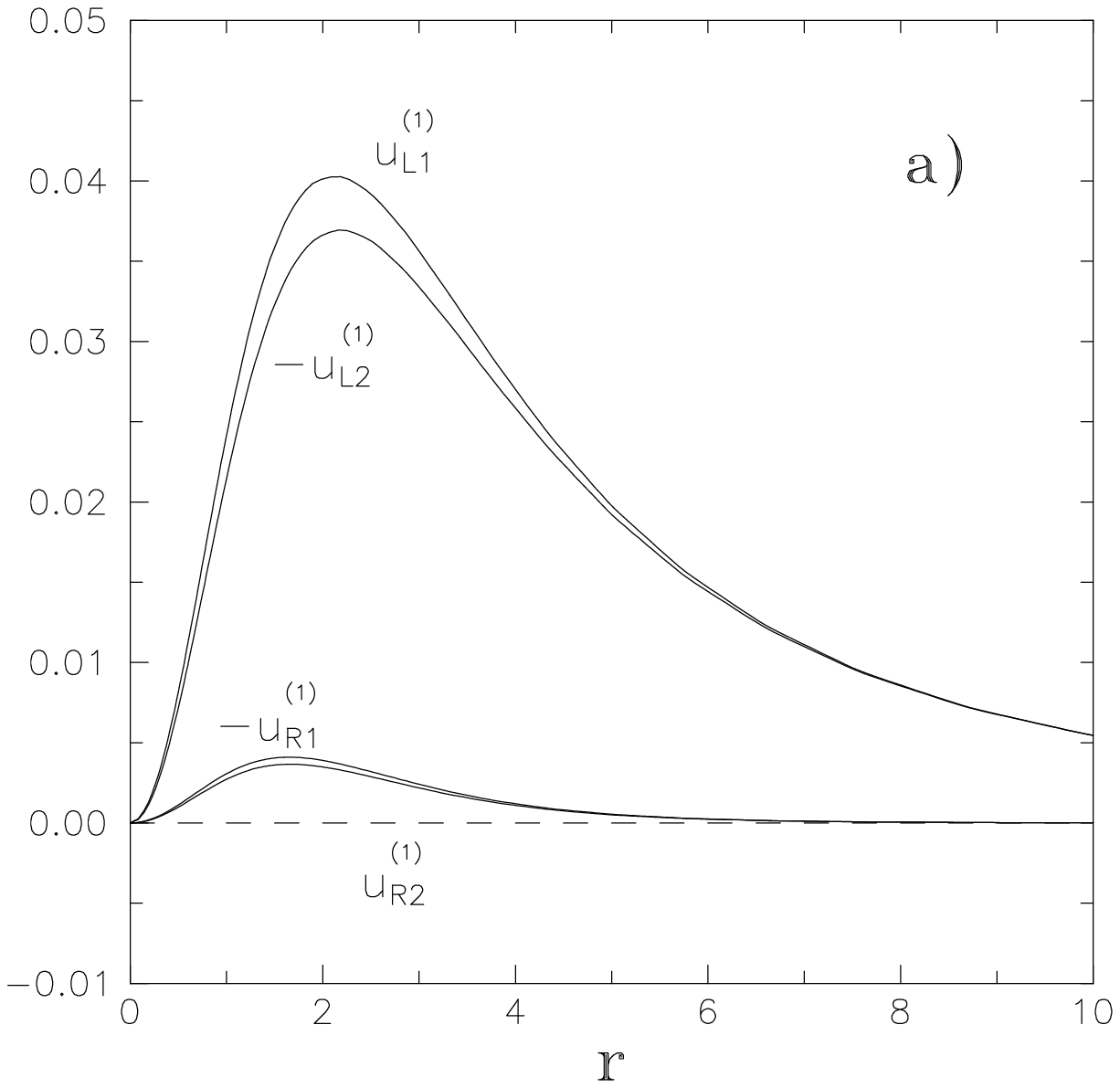,height=9cm,bbllx=4.cm,bblly=2.cm,bburx=14cm,bbury=14cm}
\psfig{figure=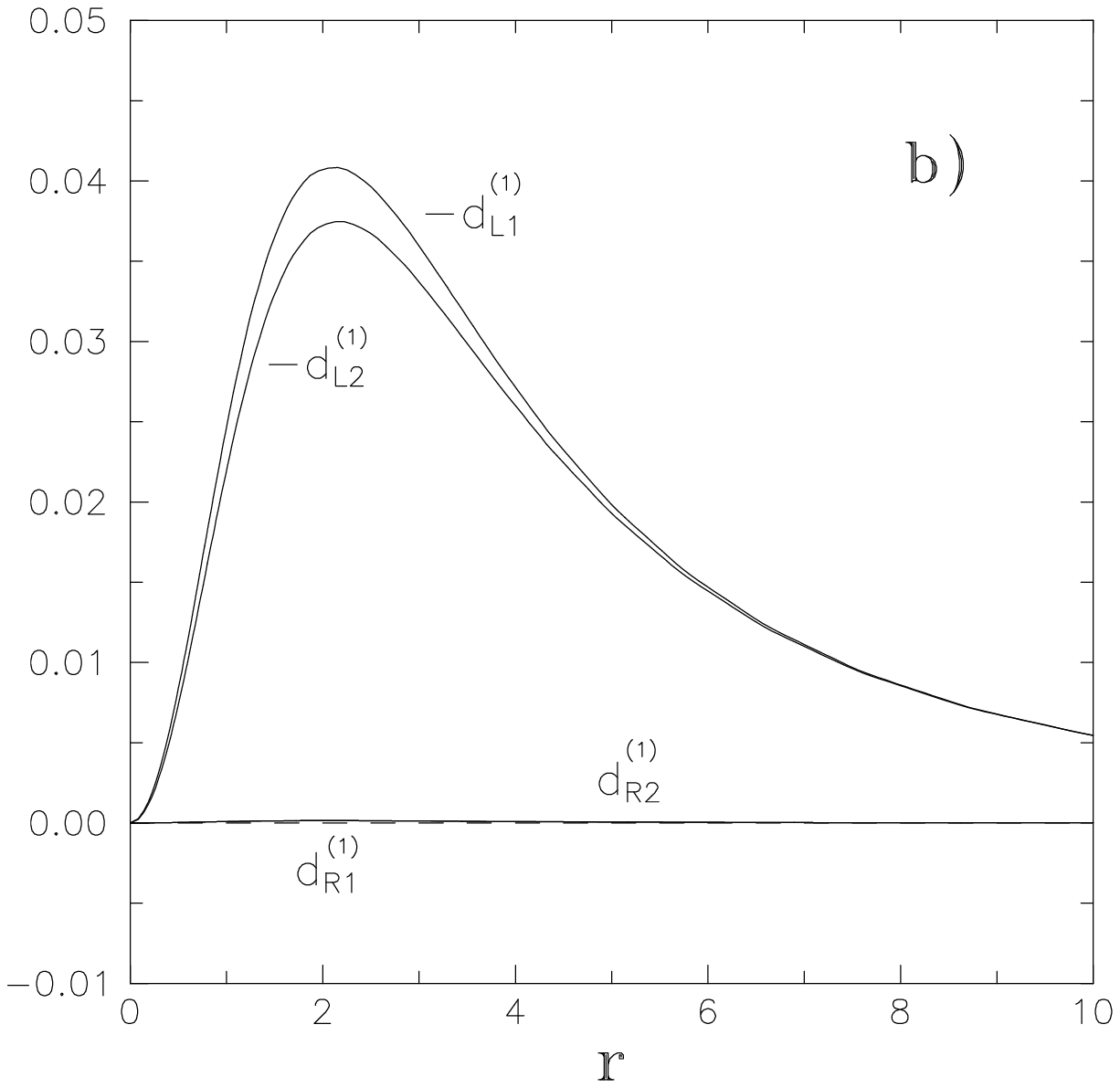,height=9cm,bbllx=4.cm,bblly=2.cm,bburx=14cm,bbury=14cm}
}}
\caption{\tenrm\baselineskip=10pt {\bf a)} The radial components
$u_{L1}^{(1)}$, $-u_{L2}^{(1)}$, $-u_{R1}^{(1)}$ and
$u_{R2}^{(1)}$ computed to order $n=1$ and $2$.
{\bf a)} The radial components
$-d_{L1}^{(1)}$, $-d_{L2}^{(1)}$, $d_{R1}^{(1)}$ and
$d_{R2}^{(1)}$ computed to order $n=1$ and $2$.}
\end{figure}
\end{document}